IAC-22-72291

# ATLAS: Deployment, Control Platform and First RSO Measurements.

João Pandeirada[a,b*], Miguel Bergano[a], Paulo Marques[c], Domingos Barbosa[a], Bruno Coelho[a], José Freitas[d], Domingos Nunes[a]

[a] *Instituto de Telecomunicações, Universidade de Aveiro, Aveiro, Portugal;* joao.pandeirada@av.it.pt (J.P); jbergano@av.it.pt (M.B); dbarbosa@av.it.pt (D.B); brunodfcoelho@av.it.pt (B.C); dfsn@av.it.pt (D.N);
[b] *Departamento de Engenharia Eletrotécnica e de Computadores, Instituto Superior Técnico, Lisboa, Portugal*
[c] *Instituto de Telecomunicações / ISEL-IPL, Lisboa, Portugal;* pmarques@isel.pt (P.M)
[d] *Ministério da Defesa Nacional, Portugal;* jose.freitas@defesa.pt (J.F)
\* Corresponding Author

**Abstract**
The ever increasing dependence of modern societies in space based services results in a rising number of objects in orbit which grows the probability of collisions between them. The increase in space debris is a threat to space assets, space based-operations and led to a common effort to develop programs for dealing with it. As part of the Portuguese Space Surveillance and Tracking (SST) project, led by the Portuguese Ministry of Defense (MoD), Instituto de Telecomunicações (IT) is developing the rAdio TeLescope pAmpilhosa Serra (ATLAS), a new monostatic radar tracking sensor located at the Pampilhosa da Serra Space Observatory (PASO), Portugal. The system operates at 5.56 GHz and aims to provide information on objects in low earth orbit (LEO), with cross sections above 10 cm2 at 1000 km. The sensor will be tasked by the Portuguese Network Operations Center (NOC), SST-PT, which interfaces with the EU-SST network. ATLAS was deployed in the first half of 2022 and is currently being tested in real case scenarios by taking range and range-rate measurements of various resident space objects (RSO). In the near future, the sensor will be completely integrated in the SST-PT network, improving the NOC services. This paper presents the deployment process of mounting the radar system in the antenna, and showcases a control platform used by operators to interact with the system. The platform control system allows remote configuration, operations scheduling and can trigger the radar system through a user friendly graphical interface. Finally, we explain the calibration procedures as well as the observation strategies we are pursuing in order to perform the initial measurements and make an accuracy assessment for ATLAS.

**Keywords:** radar; sst; space debris; leo; tracking; deployment;

**Acronyms/Abbreviations**
Low Noise Amplifier (LNA), National Operation Center (NOC), Tracking Data Message (TDM), Low Earth Orbit (LEO), Signal to Noise Ratio (SNR), Radar Cross Section (RCS), Pampilhosa da Serra Space Observatory (PASO), Application Programming Interface (API), ATLAS Cloud Service API (ACSA)

## 1. Introduction

Since the early days of the space age, humanity has been deploying various systems to orbit in the near space around Earth. This trend is likely to continue increasing as the number of space objects continues to grow and new space stakeholders enter the market [1]. Space-based services play an important role in supporting economic and social well-being, public safety, and the functioning of government responsibilities [2]. To reduce the risk of collisions in space, we need to have the ability to survey and track all the objects in our vicinity. This will provide timely and accurate information to the different stakeholders. Sensor networks monitor space assets and predict their orbits in order to avoid collisions. These networks are mainly composed of ground-based optical sensors and radar systems. The EU-SST is a support system for the creation of an autonomous monitoring network at the European level that helps to detect and track space objects, and issue alerts when necessary [3]. Portugal is a member of the EU-SST program, and is working to improve its optical and radar sensors.

IT is a third party associated with the Ministry of Defense in the EU-SST program and in 2019 initiated a major upgrade of its 9 m Cassegrain antenna located at PASO, Portugal. This upgrade is ATLAS, a monostatic ground-based radar system that can track objects up to 10 cm2 at 1000 km range [4]. ATLAS uses a solid-state power amplifier (GaN) with a peak output power of 5 kW for tracking objects in LEO. On the receiver side, it is fully coherent with detection and processing in digital domain with a bandwidth of 50 MHz and with capacity to detect Doppler velocities up to 10.79 km/s.

This articles showcases the latest developments regarding ATLAS and its operational status. In Section 2 we show the deployment on the antenna system.

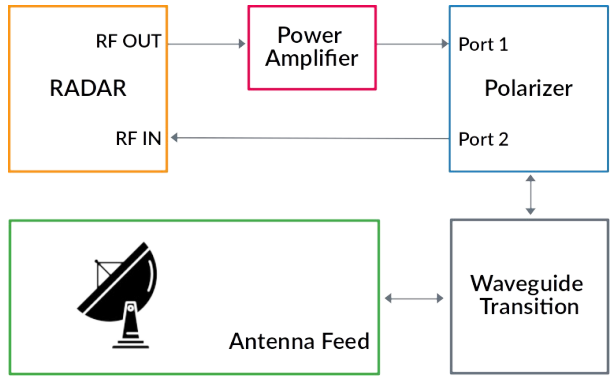

Fig. 1. Connection of the radar to the antenna feed. The polarizer is responsible of isolating the RF OUT and RF IN ports. The waveguide transition provides impedance matching between the polarizer and the feed.

Section 3 is used to present the software infrastructure and the Control Platform. Finally, in Section 4, we explain the calibration procedure and the underlying strategies in order to measure objects.

## 2. Deployment

After completing the preliminary functional tests in the lab and in the field [5], we proceeded to the deployment phase. This phase consists in the fixation of the radar system on the antenna hub and the connection of the power amplifier output to the feed.

The connection between the power amplifier output and the antenna feed is done via a polarizer and a waveguide transition as depicted in Figure 1. The polarizer is responsible for allowing the emission of only right-hand polarization EM waves as well as the amount of isolation between the output and input ports, in order to avoid damaging the sensitive LNA during transmission. Both ports should be matched to the ante-

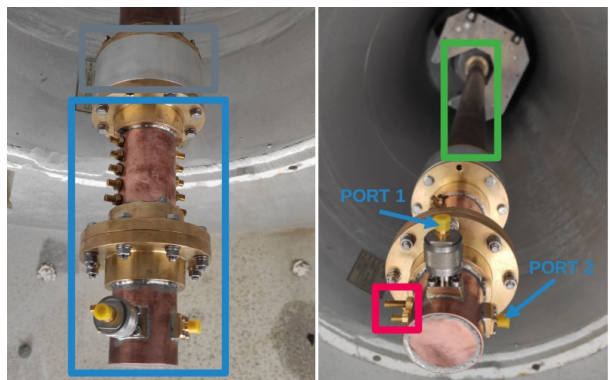

Fig. 2. Polarizer (blue box) + Waveguide Transition (grey box) connected to the antenna feed (green box). PORT 1 uses a type-N connector and is used as the emitter port. PORT 2 uses an SMA connector and is used as the receiver port. Adjusting screws (red box) are used to tune the isolation between the ports and the matching with the feed.

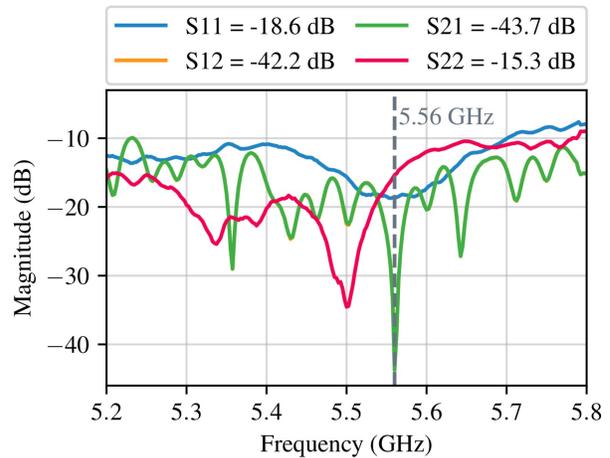

Fig. 3. S-parameters obtained after tuning. An isolation of -42.2 dB from PORT1 to PORT2 results in a maximum power leakage to the LNA during transmission of 25 dBm, which is well below the maximum allowed by the component. S11 shows a reflection coefficient of -18.6 dB which ensures that the big majority of the energy is emitted and not reflected.

-nna feed in order to maximize the amount of power emitted or received. In order to meet the requirements, the polarizer was mounted in the antenna feed (Figure 2) and the isolation and matching process was done connecting it to a VNA and tuning the S-parameters by tweaking the adjusting screws. Figure 3 plots the obtained S-parameters after the adjustment and discriminates the values for the carrier frequency of ATLAS, 5.56 GHz. A waveguide transition must be used to match the polarizer and the antenna feed.

Figure 4 shows how the radar is fixed to the antenna inside the hub. The bottom cover is a circular piece of polycarbonate which is transparent for easy visual inspection and can be removed with the appropriate screwing tools for maintenance. The hub is watertight on the upper part and has small holes in lower part for air circulation. The holes are covered with micro hole wire mesh in order to avoid the entering of external bodies inside the hub.

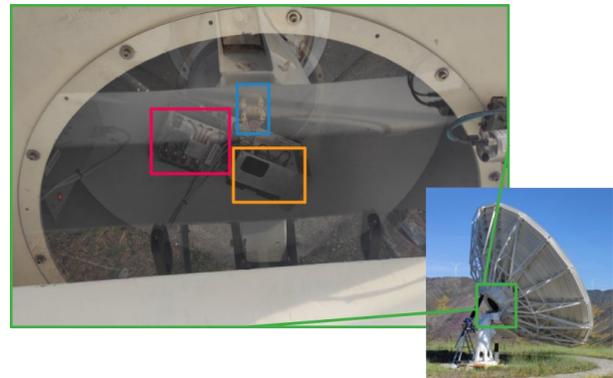

Fig. 4. Radar system installed inside the antenna hub. It is possible to see the main components of the system: radar (orange box), power amplifier (red box) and polarizer+waveguide transition (blue box).

## 3. Software Infrastructure and Control Platform

Influenced by the cloud computing concept of Everything as a Service [6], ATLAS is monitored and controlled through a web service based on an API, named ACSA. The choice of using a web based API for ATLAS is tied to the numerous advantages that this type of interface provides:
- Usage of HTTP protocol and JSON (JavaScript Object Notation) for data exchange provides ease of integration with other services that need to use ATLAS.
- Because of the separation between client and server, software development and eventual updates to the radar won't affect other services that use the system.
- Due to the long term and wide adoption of web based APIs, current development tools allow easy implementation of several sophisticated security features such as JSON schema validation, TLS encryption, token-based authentication, IP white/black listing and quotas on the number of requests.

Figure 5 shows the current software infrastructure of the sensor. The control routines of ATLAS are abstracted by ACSA that exposes an interface to the clients. ACSA works both synchronously and asynchronously depending on the type of request, this allows multiple clients to access the system in a non blocking manner. Some of the most important requests that clients can do are:
- Get status variables such as: temperatures, waveform, number of pulses, pulse repetition frequency, sampling rate, gains, delays, motor position, motor speed.
- Given a valid configuration file, configure and trigger the radar at a specified time stamp.

The raw data obtained by the radar is stored in a distributed file system. Alongside the raw I/Q data series, metadata associated with each measurement is also stored. The data can be accessed from any software system in the same network as the file system which allows parallel and remote processing of the data. Radar Signal Processing routines will feed on this file system in order to process detections and generate TDM files [7].

Since each measurement task requires a full allocation of the antenna+radar resources, the API requests are passed to a queue in order to be processed. Each client receives a task id that can be used to request the status of the task. As depicted in Figure 5, many future clients can interact with ATLAS through ACSA for different purposes, for example, a mobile application to check the status of the sensor on the go or a calibration procedure that runs periodically. The two core clients that are currently being developed are the Control Platform and the Task Scheduler.

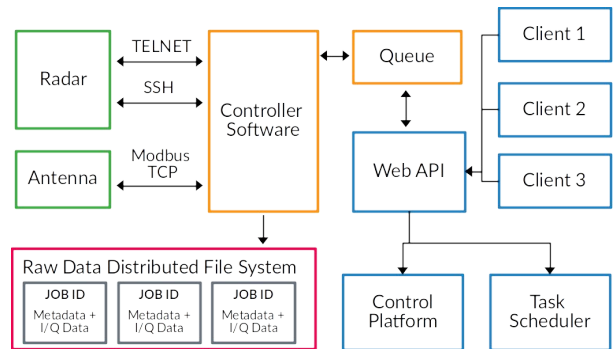

Fig. 5. The Controller Software is abstracted by ACSA and allows easy and secure access to the radar by multiple systems. Clients can only use ACSA by having a designated white-listed IP and using token-based authentication.

The Task Scheduler is the software that interfaces with the SST-PT. It is responsible for receiving tracking requests from the NOC and ensure that those tasks can be fulfilled in the most optimal way.

The Control Platform is showcased in Appendix A and allows a full picture of the whole station. It has a live video feed of the antenna surroundings, access to the status variables of the motors and the radar, a visualization of the task queue and weather variables provided by a dedicated weather station. This platform is a necessary tool for testing, monitoring, debugging and training of operators.

## 4. Calibration and Observation Strategies

Due to the narrow beamwidth of the antenna (0.73 degrees) and the high velocity of objects in LEO [4], tracking these objects is no trivial task. In order to actually hit an object in LEO orbits the sensor needs to be very accurate both in positioning and timing. Various sources of error can contribute to missing an object during observation.

Incorrect geo-referencing of the antenna homing system can introduce bias errors in the azimuth and elevation values. After appropriate geo-referencing of the homing system, a calibration using celestial sources such as the sun and the moon can be used to fine tune the offset errors of the pointing system [8]. Routines for calibration with celestial sources can be run weekly/monthly to maintain up to date values of the offsets of the antenna. Errors in orbit prediction can cause both errors in the state vectors as well as in the time stamps [9].

An important step to mitigate some of these sources of error is the choice of objects to track in the calibration phase. As an example, the ISS can be used for initial measurements since it has a very large RCS (around 400 m$^2$) and can orbit in the low LEO region (below 500 km of altitude), making it possible to receive higher energy echoes in comparison with other RSOs. Moreover, the ISS orbit is well defined and constantly adjusted, minimizing the errors associated with its orbit prediction. Other LEO operational satellites with telemetry and retro reflectors such as the CRYOSAT-2 and Jason-3 offer precise ephemeris information in the form of CPF files [10] that can be used to predict their passages with high accuracy, providing valuable information for both calibration procedures as well as performance assessment of ground based sensors [11].

Passive calibration satellites are objects with known and constant RCS. Such objects like the STELLA and STARLETTE provide high reflectivity and are spherical in order to provide an RCS that does not depend on the aspect angle [12]. These objects are used to tune the RCS measurements obtained by the sensors since they provide a ground truth for the RCS.

The observation strategy used can also mitigate some of the errors mentioned before. As for the first measurements, we propose a semi-surveillance scheme. This scheme consists in parking the antenna at a predicted position for the satellite of interest illuminating that region of the sky some time before and after the predicted time for when the object passes inside the beamwidth. This method will increase the probability of hitting the target and the echoes will provide useful information on how to compensate deviations in the orbit prediction.

## 5. Conclusions and Future Work

After completion of all the laboratory experiments, ATLAS was finally connected to the antenna located at PASO. The deployment consisted in connecting the output of the power amplifier to the antenna feed through a polarizer. The polarizer passed through a matching procedure in order to maximize the isolation between the input/output ports and minimize reflections. The system is safely fixed to the antenna and images of the setup were provided.

The radar follows a sensor as a service philosophy where multiple clients can access the system in a non blocking manner in a reliable and secure way. A Control Platform is showcased where the radar can be monitored and triggered by operators.

The task of tracking objects in LEO orbits demands a complex calibration process and a careful observation plan. Several sources of error are listed and methods for mitigating them are presented. As of now we are implementing the proposed strategies and in the near future we will present measurements and an accuracy assessment to evaluate the performance of the sensor as well as next steps to improve it.

## Acknowledgements
This work is supported by the Fundação para Ciência e Tecnologia (FCT), Ph.D. grant No.2022.12341.BDANA. This paper and team work has been supported by the European Commission H2020 Programme under the grant agreement 2-3SST2018-20; The team acknowledges LC Technologies for technical developments in close collaboration with IT and ATLAR for helpful comments. The team acknowledges financial support from ENGAGE-SKA Research Infrastructure, ref. POCI-01-0145-FEDER-022217, funded by COMPETE 2020 and FCT, Portugal; IT team members acknowledge support from Projecto Lab. Associado UID/EEA/50008/2019.

### Appendix A (Control Platform User Interface)

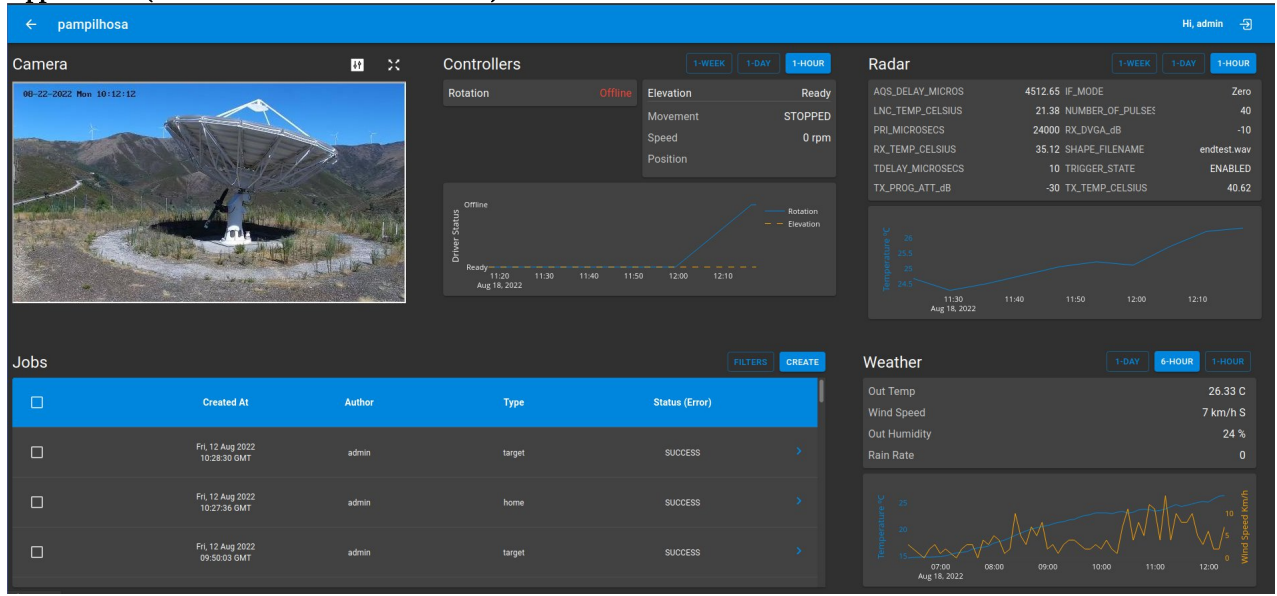